\documentstyle[12pt,epsfig,graphics,cite]{article}
\begin{document}\bibliographystyle{plain}\begin{titlepage}
\renewcommand{\thefootnote}{\fnsymbol{footnote}}\hfill\begin{tabular}{l}
HEPHY-PUB 812/05\\hep-ph/0510372\\October 2005\end{tabular}\\[1cm]
\Large\begin{center}{\bf EXACT-PROPAGATOR INSTANTANEOUS BETHE--SALPETER
EQUATION FOR QUARK--ANTIQUARK BOUND STATES}\\[1.5cm]\large{\bf LI
Zhi-Feng}\\[.3cm]\normalsize Institut f\"ur Theoretische Physik,
Universit\"at Wien,\\Boltzmanngasse 5, A-1090 Wien, Austria\\[1cm]\large{\bf
Wolfgang LUCHA\footnote[1]{\normalsize\ {\em E-mail address\/}:
wolfgang.lucha@oeaw.ac.at}}\\[.3cm]\normalsize Institut f\"ur
Hochenergiephysik,\\\"Osterreichische Akademie der
Wissenschaften,\\Nikolsdorfergasse 18, A-1050 Wien, Austria\\[1cm]\large{\bf
Franz F.~SCH\"OBERL\footnote[2]{\normalsize\ {\em E-mail address\/}:
franz.schoeberl@univie.ac.at}}\\[.3cm]\normalsize Institut f\"ur Theoretische
Physik, Universit\"at Wien,\\Boltzmanngasse 5, A-1090 Wien,
Austria\vfill{\normalsize\bf Abstract}\end{center}\normalsize Recently an
instantaneous approximation to the Bethe--Salpeter formalism for the analysis
of bound states in quantum field theory has been proposed which retains, in
contrast to the Salpeter equation, as far as possible the exact propagators
of the bound-state constituents, extracted nonperturbatively from
Dyson--Schwinger equations or lattice gauge theory. The implications of this
improvement for the solutions of this bound-state equation, that is, the
spectrum of the mass eigenvalues of its bound states and the corresponding
wave~functions, when considering the quark propagators arising in quantum
chromodynamics are explored.\vspace{.5cm} {\em PACS numbers\/}: 11.10.St,
03.65.Ge, 03.65.Pm
\renewcommand{\thefootnote}{\arabic{footnote}}\end{titlepage}

\section{Introduction}The more than half a century old Bethe--Salpeter
formalism\cite{BSE} constitutes a relativistically covariant framework within
the realms of quantum field theory for the description of bound states from
first principles. The Bethe--Salpeter equation controls a bound-state
amplitude encoding, in momentum space, the distribution of the relative
momenta of the bound-state constituents. Within elementary particle physics
this formalism has been widely applied to quantum electrodynamics (QED) and
quantum chromodynamics (QCD). Unfortunately it faces problems of
interpretation and of our ignorance of the full interaction kernel~in~QCD.

Thus, simplifications of the Bethe--Salpeter equation in form of some
three-dimensional reductions are highly desirable. The most famous among all
proposals is known as Salpeter equation\cite{SE}. Its formulation, however,
is based on assuming all bound-state constituents to interact instantaneously
{\em and\/} to propagate as free particles; the latter circumstance~renders
hard to implement effects such as spontaneous chiral-symmetry breaking,
crucial for QCD.

In view of this, an instantaneous Bethe--Salpeter equation which incorporates
the exact form of the propagators of the bound-state constituents (to the
utmost conceivable extent) has been derived recently\cite{Lucha05:IBSEWEP};
this improved bound-state equation reduces, of course, to the Salpeter
equation, upon approximation of the exact propagators by their free
counterparts. For any description of hadrons as bound states, the exact quark
propagators conforming to the QCD Dyson--Schwinger equations are relevant.
This work is devoted to the study~of~the consequences of introducing exact
quark propagators in this instantaneous~Bethe--Salpeter equation. The most
dramatic effect observed is a significant diminution of the level spacing.

The paper is organized as follows. Section~\ref{Sec:IBSEWEP} sketches the
derivation of our instantaneous Bethe--Salpeter equation with exact
propagators of the bound-state constituents presented in
Ref.\cite{Lucha05:IBSEWEP}. Approximating all interactions entering in the
Bethe--Salpeter equation by their static forms but retaining, as far as
possible, exact propagators yields\cite{Lucha05:IBSEWEP} a generalization of
Salpeter's equation\cite{SE}, with momentum-dependent masses of the
bound-state constituents and with normalization factors of their exact
propagators multiplying all interaction terms. Section~\ref{Sec:QP}
introduces the exact light-quark propagators obtained within QCD as solution
of the Dyson--Schwinger equations. This infinite tower of coupled integral
equations calls for a truncation\cite{Maris97a,Maris97b,Roberts98,Ivanov98,
Maris99a,Maris99b,Roberts00a,Roberts00b,Alkofer00,Maris00,Maris01,Maris02,
Bhagwat02,Tandy03,Maris03,Bhagwat03,Roberts03,Krassnigg03a,Krassnigg03b,
Alkofer03a,Krassnigg04} which must not be in conflict with the relevant
Ward--Takahashi identity. Section~\ref{Sec:PSFAFBS} summarizes the technique
developed for finding the solutions of an instantaneous Bethe--Salpeter
equation by first reducing it to a coupled set of radial
equations\cite{Lagae92a,Lagae92b,Olsson95,Olsson96}~and then converting it to
a matrix eigenvalue problem\cite{Lucha00:IBSEm=0,Lucha00:IBSE-C4,
Lucha00:IBSEnzm,Lucha01:IBSEIAS}. Some implications of taking into account
exact instead of free propagators of bound-state constituents are
analyzed~in~Sec.~\ref{Sec:RD} by application of the entire formalism to a
linear confining interaction. Section~\ref{Sec:SCO} scrutinizes our findings.
Appendix~\ref{App:GLB} recalls the Hilbert-space basis required for the
matrix conversion.

\section{Instantaneous Bethe--Salpeter equation with exact propagators}
\label{Sec:IBSEWEP}The derivation of the {\em exact-propagator\/}
instantaneous Bethe--Salpeter equation\cite{Lucha05:IBSEWEP} parallels the
(three-dimensional) reduction of the Bethe--Salpeter equation\cite{BSE} to
the {\em free-propagator\/} Salpeter equation\cite{SE}. It may be achieved by
several slightly different but equivalent routes.

In the framework of the Bethe--Salpeter formalism, a bound state $|{\rm
B}(P)\rangle$ of momentum $P$ and mass $M_{\rm B},$ composed of a fermion and
an antifermion described by the field operators $\psi_1(x_1),$
$\bar\psi_2(x_2),$ resp., is represented, in momentum space, by the
Bethe--Salpeter amplitude$$\Psi(p)\equiv\exp({\rm i}\,P\,X)\int{\rm
d}^4x\,\exp({\rm i}\,p\,x)\,\langle 0|{\rm T}(\psi_1(x_1)\,\bar\psi_2(x_2))
|{\rm B}(P)\rangle\ .$$Here, $X$ and $x$ denote the center-of-momentum and
relative coordinates of the two-particle system while $P$ and $p$ label the
total and relative momenta of the bound-state constituents. This
Bethe--Salpeter amplitude $\Psi$ has to satisfy the homogeneous
Bethe--Salpeter equation\begin{equation}\Psi(p)=\frac{{\rm
i}}{(2\pi)^4}\,S_1(p_1)\int{\rm d}^4q\,K(p,q)\,\Psi(q)\,S_2(-p_2)\
.\label{Eq:BSE}\end{equation}The dynamical ingredients of this equation of
motion are the exact propagators $S_i(p)$~of~the two bound fermions $i=1,2$
(with individual momenta $p_1,$ $p_2$) and the interaction kernel $K,$ a
fully truncated 4-point Green function which encompasses all Bethe--Salpeter
irreducible Feynman diagrams for two-particle into two-particle scattering
and depends on the relative momenta of initial and final scattering states,
$p$ and $q,$ as well as on the total momentum~$P$.

The instantaneous approximation to this formalism assumes that the kernel $K$
depends just on the spatial components $\mbox{\boldmath{$p$}}$ and
$\mbox{\boldmath{$q$}}$ of the relative momenta $p$ and $q$: $K(p,q)=\hat
K(\mbox{\boldmath{$p$}},\mbox{\boldmath{$q$}}).$ Its application reduces
Eq.~(\ref{Eq:BSE}) to the instantaneous version of the Bethe--Salpeter
equation\begin{equation}\Phi(\mbox{\boldmath{$p$}})=\frac{{\rm i}}{2\pi}
\int{\rm d}p_0\,S_1(p_1)\,I(\mbox{\boldmath{$p$}})\,S_2(-p_2)\label{Eq:BSE-I}
\end{equation}for the Salpeter amplitude (defined by integration of $\Psi(p)$
over the time component $p_0$~of~$p$)$$\Phi(\mbox{\boldmath{$p$}})\equiv
\frac{1}{2\pi}\int{\rm d}p_0\,\Psi(p)\ ;$$ here the term involving the by
assumption now instantaneous interaction is abbreviated~by
$$I(\mbox{\boldmath{$p$}})\equiv\frac{1}{(2\pi)^3}\int{\rm d}^3q\,\hat
K(\mbox{\boldmath{$p$}},\mbox{\boldmath{$q$}})\, \Phi(\mbox{\boldmath{$q$}})\
.$$

The fermion propagator $S_i(p)$ is the solution of the fermion
Dyson--Schwinger equation. By Lorentz covariance $S_i(p)$ is defined by
merely two (Lorentz-scalar) functions $M_i(p^2)$~and $Z_i(p^2),$ in QCD
referred to as the quark wave-function renormalization and mass functions:
$$S_i(p)=\frac{{\rm i}\,Z_i(p^2)}{\not\!p-M_i(p^2)+{\rm i}\,\varepsilon}\
,\quad\not\!p\equiv p^\mu\,\gamma_\mu\ .$$In the course of the derivation
\cite{Lucha05:IBSEWEP} of a generalization of the Salpeter equation
towards~exact propagators of all bound-state constituents, two of the present
authors (W.~L. and~F.~F.~S.) assumed these functions, $M_i(p^2)$ and
$Z_i(p^2),$ to depend only on the spatial components $\mbox{\boldmath{$p$}}$
of the momentum $p.$ This allows to substitute $M_i(p^2)$ by
$M_i(\mbox{\boldmath{$p$}}^2)$ and $Z_i(p^2)$ by
$Z_i(\mbox{\boldmath{$p$}}^2)$~in~$S_i(p).$

Then the integral in Eq.~(\ref{Eq:BSE-I}) over the time component $p_0$ can
be easily given analytically. Introducing the one-particle energy
$E_i(\mbox{\boldmath{$p$}}),$ the (generalized) Dirac Hamiltonian
$H_i(\mbox{\boldmath{$p$}}),$ and the energy projection operators
$\Lambda_i^\pm(\mbox{\boldmath{$p$}})$ for positive or negative energy of
particle $i=1,2$~by\begin{eqnarray*}E_i(\mbox{\boldmath{$p$}})&\equiv&
\sqrt{\mbox{\boldmath{$p$}}^2+M_i^2(\mbox{\boldmath{$p$}}^2)}\ ,\quad i=1,2\
,\\[1ex]H_i(\mbox{\boldmath{$p$}})&\equiv&
\gamma_0\,[\mbox{\boldmath{$\gamma$}}\cdot\mbox{\boldmath{$p$}}
+M_i(\mbox{\boldmath{$p$}}^2)]\ ,\quad i=1,2\ ,\\[1ex]
\Lambda_i^\pm(\mbox{\boldmath{$p$}})&\equiv&\frac{E_i(\mbox{\boldmath{$p$}})\pm
H_i(\mbox{\boldmath{$p$}})}{2\,E_i(\mbox{\boldmath{$p$}})}\ ,\quad i=1,2\
,\end{eqnarray*}few rather standard manipulations yield\cite{Lucha05:IBSEWEP}
our instantaneous Bethe--Salpeter equation for fermion--antifermion bound
states, with exact propagators of the bound-state
constituents:\begin{equation}\Phi(\mbox{\boldmath{$p$}})
=Z_1(\mbox{\boldmath{$p$}}_1^2)\,Z_2(\mbox{\boldmath{$p$}}_2^2)
\left(\frac{\Lambda_1^+(\mbox{\boldmath{$p$}}_1)\,\gamma_0\,
I(\mbox{\boldmath{$p$}})\,\gamma_0\,\Lambda_2^-(\mbox{\boldmath{$p$}}_2)}
{P_0-E_1(\mbox{\boldmath{$p$}}_1)-E_2(\mbox{\boldmath{$p$}}_2)}
-\frac{\Lambda_1^-(\mbox{\boldmath{$p$}}_1)\,\gamma_0\,
I(\mbox{\boldmath{$p$}})\,\gamma_0\,\Lambda_2^+(\mbox{\boldmath{$p$}}_2)}
{P_0+E_1(\mbox{\boldmath{$p$}}_1)+E_2(\mbox{\boldmath{$p$}}_2)}\right).
\label{Eq:IBSE}\end{equation}From this, each amplitude
$\Phi(\mbox{\boldmath{$p$}})$ has to satisfy the two constraints
$\Lambda_1^\pm(\mbox{\boldmath{$p$}}_1)\,\Phi(\mbox{\boldmath{$p$}})\,
\Lambda_2^\pm(\mbox{\boldmath{$p$}}_2)=0.$

With little effort our bound-state equation (\ref{Eq:IBSE}) may be rephrased
as eigenvalue problem,
\begin{eqnarray*}&&H_1(\mbox{\boldmath{$p$}}_1)\,\Phi(\mbox{\boldmath{$p$}})
-\Phi(\mbox{\boldmath{$p$}})\,H_2(\mbox{\boldmath{$p$}}_2)\\[1ex]
&&+\,Z_1(\mbox{\boldmath{$p$}}_1^2)\,Z_2(\mbox{\boldmath{$p$}}_2^2)\,
[\Lambda_1^+(\mbox{\boldmath{$p$}}_1)\,\gamma_0\,I(\mbox{\boldmath{$p$}})\,
\gamma_0\,\Lambda_2^-(\mbox{\boldmath{$p$}}_2)
-\Lambda_1^-(\mbox{\boldmath{$p$}}_1)\,\gamma_0\,I(\mbox{\boldmath{$p$}})\,
\gamma_0\,\Lambda_2^+(\mbox{\boldmath{$p$}}_2)]\\[1ex]
&&=P_0\,\Phi(\mbox{\boldmath{$p$}})\ ,\end{eqnarray*}with the bound state's
energy $P_0$ or, in its rest frame
$\mbox{\boldmath{$p$}}_2=-\mbox{\boldmath{$p$}}_1,$ the mass $M_{\rm B}$ as
eigenvalue.

The Salpeter equation\cite{SE} is obtained by one further step of
simplification. Its derivation assumes, in addition to the instantaneous
approximation for $K$, that each exact propagator in Eq.~(\ref{Eq:BSE}) can
be replaced by the propagator $S_0(p,m_i)$ of a free particle of effective
mass~$m_i$:$$S_i(p)\cong S_0(p,m_i)=\frac{{\rm i}}{\not\!p-m_i+{\rm
i}\,\varepsilon}\equiv{\rm i}\,\frac{\not\!p+m_i}{p^2-m_i^2+{\rm
i}\,\varepsilon}\ ,\quad i=1,2\ .$$Thus the Salpeter equation is recovered
from Eq.~(\ref{Eq:IBSE}) in the limit $M_i(p^2)\to m_i,$ $Z_i(p^2)\to 1.$ The
exact-propagator instantaneous Bethe--Salpeter equation (\ref{Eq:IBSE})
generalizes\cite{Lucha05:IBSEWEP} Salpeter's equation by replacing $m_i$ by
$M_i(p^2)$ and introducing factors $Z_i(p^2)$ in the interaction terms.

\section{Quark propagator from Dyson--Schwinger equation}\label{Sec:QP}Within
QCD the Dyson--Schwinger equation for the quark propagator involves, besides
the exact quark propagator, also the exact gluon propagator and the exact
quark--gluon vertex. The Dyson--Schwinger equations governing the two latter
Green functions couple the quark Dyson--Schwinger equation to the infinite
hierarchy of Dyson--Schwinger equations.~Hence, a tractable problem can only
be defined by some truncation of this set of integral equations.

For the present investigation we employ the so-called
``renormalization-group-improved rainbow--ladder truncation'' scheme
\cite{Maris97a,Maris97b,Roberts98,Ivanov98,Maris99a,Maris99b,Roberts00a,
Roberts00b,Alkofer00,Maris00,Maris01,Maris02,Bhagwat02,Tandy03,Maris03,
Bhagwat03,Roberts03,Krassnigg03a,Krassnigg03b,Alkofer03a,Krassnigg04} applied
to the quark Dyson--Schwinger equation and the meson Bethe--Salpeter
equation. In this specific scheme the exact gluon propagator and the exact
quark--gluon vertex are replaced by their (perturbative) tree-level
forms.~The truncation is consistent with the preservation of the axial-vector
Ward--Takahashi identity; this is important for all questions related to
chiral symmetry and its dynamical breakdown. In this model, all the dynamical
information is encoded in some effective coupling strength.

Viewed as function of the involved momentum transfer squared this effective
coupling is characterized by two main features. In the ultraviolet region it
approaches the perturbative behaviour of the strong fine-structure constant,
incorporating thereby asymptotic freedom. In the infrared region it exhibits
the significant enhancement demanded strongly by studies of the
Dyson--Schwinger equation satisfied by the exact gluon propagator. In the
particular {\em Ansatz\/} for this effective coupling strength proposed in
Ref.\cite{Maris97a} this infrared enhancement is represented partly by the
integrable singularity of a momentum-space $\delta$ function, partly by a
finite-width approximation to this $\delta$ function. This constitutes the
``model of our~choice.''

\newpage

In the comprehensive study presented in Ref.\cite{Maris97a}, the exact quark
propagators emerging from this truncation model are found as numerical
solutions of the quark Dyson--Schwinger equation by fitting main properties
of the $\pi$- and {\sl K}-meson system. Like many treatments of
Dyson--Schwinger equations the analysis of Ref.\cite{Maris97a} has been
performed in Euclidean space, implying that the quark propagators are
obtained as Euclidean-space Schwinger functions. Within both QED and QCD, the
analytic structure of the exact fermion propagators is still the subject of
intense investigations which did not provide a definitive conclusion until
now (cf., for instance, Refs.\cite{Maris94,Alkofer03a,Alkofer03b} and
references therein). In order to proceed, we must thus assume that the
necessary analytic continuation from Euclidean to Minkowski space makes
sense, at least for the quark propagators. In this case, the numerically
computed functions $M(p^2)$ and $Z(p^2)$ in the propagator of the light {\sl
u}- and {\sl d}-quarks may be represented (rather accurately in a range of
spacelike momenta) in analytical form by the parametrizations
\cite{Roberts02:PC}$$M(p^2)=\frac{a}{1+\displaystyle\frac{p^4}{b}}+m_0\
,\quad Z(p^2)=1-\frac{c}{1-\displaystyle\frac{p^2}{d}}\ ;$$the values of the
parameters $a,b,c,d,m_0$ are fixed by interpolation of the numerical
results:\begin{eqnarray}&&a=0.745\;{\rm GeV}\ ,\quad b=(0.744\;{\rm GeV})^4\
,\quad m_0=0.0055\;{\rm GeV}\ ,\nonumber\\[1ex]&&c=0.545\ ,\quad
d=(1.85508\;{\rm GeV})^2\ .\label{Eq:ParVal}\end{eqnarray}These propagator
functions read in their ``$p_0^2=0$'' approximation required for an
analytical formulation of the bound-state equation (\ref{Eq:IBSE}) proposed
\cite{Lucha05:IBSEWEP} as a generalized Salpeter equation\begin{equation}
M(\mbox{\boldmath{$p$}}^2)=
\frac{a}{1+\displaystyle\frac{\mbox{\boldmath{$p$}}^4}{b}}+m_0\ ,\quad
Z(\mbox{\boldmath{$p$}}^2)=1-
\frac{c}{1+\displaystyle\frac{\mbox{\boldmath{$p$}}^2}{d}}\
.\label{Eq:ProPar}\end{equation}

\begin{figure}[p]\begin{center}\begin{tabular}{c}
\psfig{figure=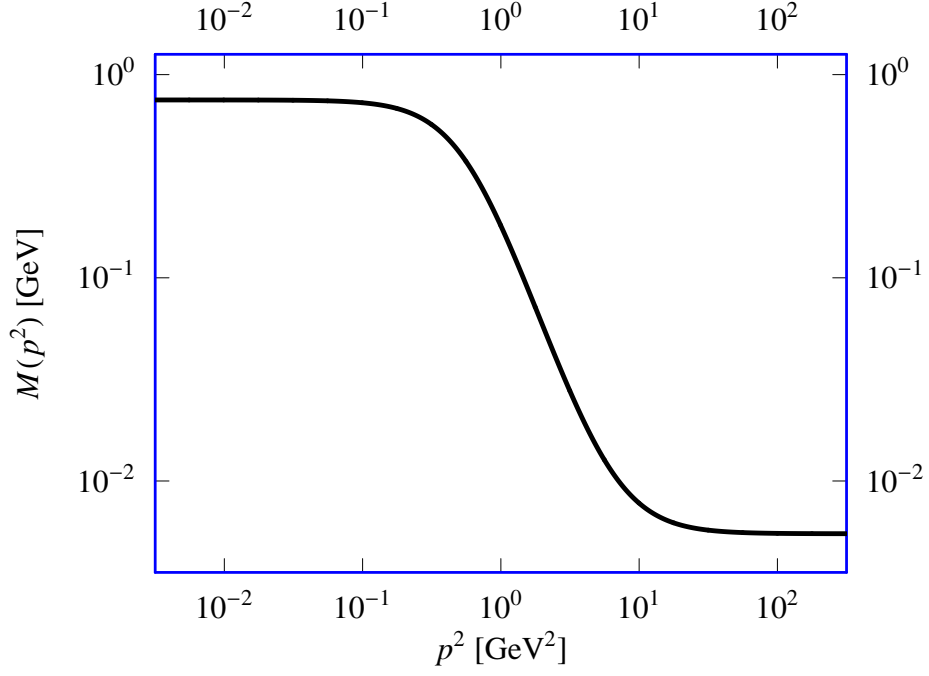,scale=1}\\[0.05ex](a)\end{tabular}\\[1.5ex]
\begin{tabular}{c}\psfig{figure=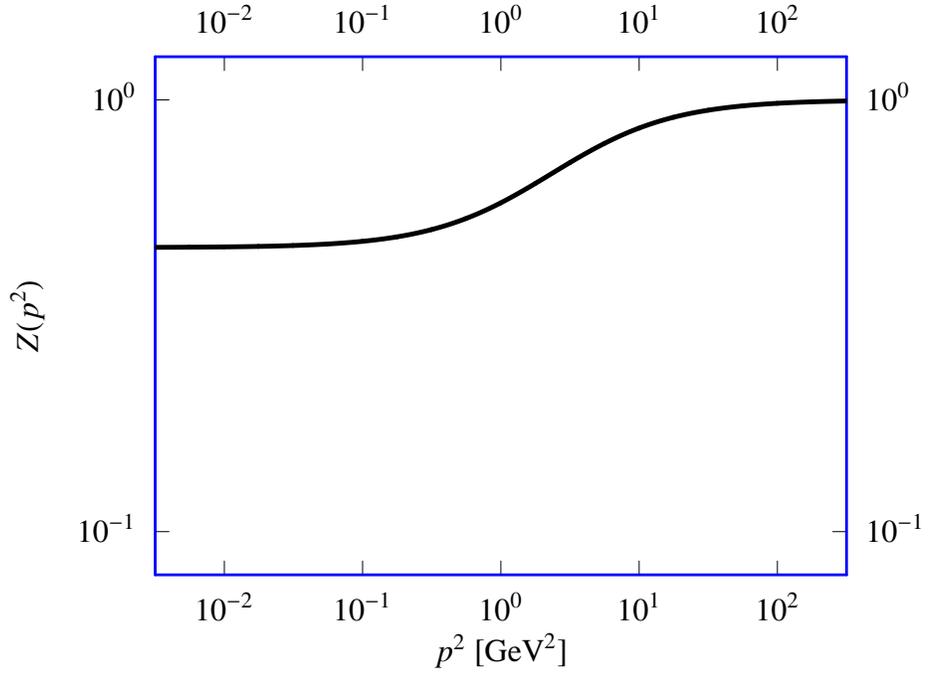,scale=1}\\[0.05ex](b)
\end{tabular}\\[0.5ex]\caption{Mass function $M(\mbox{\boldmath{$p$}}^2)$ (a)
and wave-function renormalization function $Z(\mbox{\boldmath{$p$}}^2)$ (b)
of the exact propagator of the (light) {\sl u}- and {\sl d}-quarks obtained
by numerical solution of the quark Dyson--Schwinger equation in the
``renormalization-group-improved rainbow--ladder truncation'' model of
Ref.\cite{Maris97a} as represented by the simple interpolation of
Eqs.~(\ref{Eq:ProPar})~and~(\ref{Eq:ParVal}).}\label{Fig:PF}\end{center}
\end{figure}

The behaviour of the quark propagator functions $M(\mbox{\boldmath{$p$}}^2)$
and $Z(\mbox{\boldmath{$p$}}^2)$ as functions of $\mbox{\boldmath{$p$}}^2$ is
depicted in Fig.~\ref{Fig:PF}. For light quarks the mass function
$M(\mbox{\boldmath{$p$}}^2)$ is, of course, dominated by the nonperturbative
mechanism responsible for dynamical chiral-symmetry breaking.~Starting at
$M(0)=0.7505\;{\rm GeV},$ $M(\mbox{\boldmath{$p$}}^2)$ drops in the vicinity
of $\mbox{\boldmath{$p$}}^2=(0.57\;{\rm GeV})^2$ by more than two orders of
magnitude, in order to approach in the limit
$\mbox{\boldmath{$p$}}^2\to\infty$ the (comparatively~tiny but still
nonvanishing and hence explicitly chiral-symmetry breaking) current
light-quark mass $m_0=0.0055\;{\rm GeV}.$ In contrast to such drastic
variation, the wave-function renormalization function
$Z(\mbox{\boldmath{$p$}}^2)$ exhibits an only rather moderate dependence on
$\mbox{\boldmath{$p$}}^2.$ With increasing values of
$\mbox{\boldmath{$p$}}^2,$ $Z(\mbox{\boldmath{$p$}}^2)$ rises slowly from
$Z(0)=0.455$ to its asymptotic value $Z(\mbox{\boldmath{$p$}}^2)\to1$ for
$\mbox{\boldmath{$p$}}^2\to\infty.$

Within the ``renormalization-group-improved rainbow--ladder truncation''
scheme, it is by no means mandatory to implement, as done in the model
studied in Refs.\cite{Maris97a,Maris97b,Roberts98,Ivanov98,Roberts00a,
Roberts00b,Krassnigg04}, the infrared enhancement in the effective
interaction by the sum of an integrable $\delta$ function singularity and its
finite-width approximation. The results of the investigations reported in
Refs.\cite{Maris99a,Maris99b,Maris00,Maris01,Maris02,Bhagwat02,Tandy03,Maris03,
Roberts03,Alkofer03a} demonstrate that propagator functions of very similar
shape will be obtained in a model in which the infrared enhancement of the
effective-interaction coupling required by hadron phenomenology is provided
only by the finite-width representation\cite{Alkofer00}.

Moreover, the predictions for the propagator functions $M(p^2)$ and $Z(p^2)$
of both models\cite{Maris97a,Maris99a} for the effective coupling in the
quark Dyson--Schwinger equation exhibit a remarkable qualitative and
quantitative agreement with the results produced by lattice gauge theories. A
recent {\em unquenched\/} lattice calculation of the quark propagator in
Landau gauge involving two degenerate light ({\sl u\/}/{\sl d\/}) and one
heavier ({\sl s\/}) dynamical quarks may be found in Ref.\cite{Bowman05}.

\section{Pseudoscalar fermion--antifermion bound states}\label{Sec:PSFAFBS}
Now follow the path, paved in Refs.\cite{Lagae92a,Lagae92b,Olsson95,Olsson96,
Lucha00:IBSEm=0,Lucha00:IBSE-C4,Lucha00:IBSEnzm,Lucha01:IBSEIAS}, of the
transformation of bound-state equations for Salpeter amplitudes
$\Phi(\mbox{\boldmath{$p$}})$ to matrix eigenvalue problems fixing their
radial components.

Consider, as the perhaps simplest example, fermion--antifermion bound states
of spin $J,$ parity $P=(-1)^{J+1}$ and charge-conjugation quantum number
$C=(-1)^J.$ In spectroscopic notation such states are labeled by ${}^1J_J.$
Because of the constraints $\Lambda_1^\pm(\mbox{\boldmath{$p$}}_1)\,
\Phi(\mbox{\boldmath{$p$}})\,\Lambda_2^\pm(\mbox{\boldmath{$p$}}_2)=0$ the
general expansion of the Salpeter amplitude $\Phi(\mbox{\boldmath{$p$}})$
over a complete set of Dirac matrices involves not the full 16 but only eight
independent components. For our ${}^1J_J$ states only two of the latter,
$\Phi_1(\mbox{\boldmath{$p$}})$ and $\Phi_2(\mbox{\boldmath{$p$}}),$ are
relevant. With our notation for one-particle energy
$E(\mbox{\boldmath{$p$}})$ and (generalized) Dirac Hamiltonian
$H(\mbox{\boldmath{$p$}})$ introduced in Sec.~\ref{Sec:IBSEWEP} the
corresponding Salpeter amplitude $\Phi(\mbox{\boldmath{$p$}})$ reads in the
center-of-momentum frame of the particle--antiparticle
system$$\Phi(\mbox{\boldmath{$p$}})=\left[\Phi_1(\mbox{\boldmath{$p$}})\,
\frac{H(\mbox{\boldmath{$p$}})}{E(\mbox{\boldmath{$p$}})}
+\Phi_2(\mbox{\boldmath{$p$}})\right]\gamma_5\ .$$Without loss of generality
but for definiteness, focus to exactly the same physical system as studied in
Refs.\cite{Lucha00:IBSEm=0,Lucha00:IBSE-C4,Lucha00:IBSEnzm,Lucha01:IBSEIAS}.
Aiming at the description of mesons with the quantum numbers of the pion
discuss quark--antiquark bound states of spin $J=0,$ that is, pseudoscalar
states~of spin-parity-charge conjugation assignment $J^{PC}=0^{-+}.$ Assume
the kernel $\hat K(\mbox{\boldmath{$p$}},\mbox{\boldmath{$q$}})$ to be of
convolution type with Dirac structure that of time-component Lorentz-vector
interactions: $\hat K(\mbox{\boldmath{$p$}},\mbox{\boldmath{$q$}})=\hat
K(\mbox{\boldmath{$p$}}-\mbox{\boldmath{$q$}})=
V(\mbox{\boldmath{$p$}}-\mbox{\boldmath{$q$}})\,\gamma^0\otimes\gamma^0$ with
$V(\mbox{\boldmath{$p$}}-\mbox{\boldmath{$q$}})$ any Lorentz-scalar potential
function.

Upon factorizing off all dependence of the Salpeter amplitude
$\Phi(\mbox{\boldmath{$p$}})$ on angular variables the exact-propagator
instantaneous Bethe--Salpeter equation (\ref{Eq:IBSE}) may be reduced
\cite{Lagae92a,Olsson95} to a set of coupled equations for the radial factors
of all the independent Salpeter components. For bound states composed of
particle and corresponding antiparticle we clearly have, with
$p\equiv|\mbox{\boldmath{$p$}}|,$
$Z_1(\mbox{\boldmath{$p$}}^2)=Z_2(\mbox{\boldmath{$p$}}^2)\equiv Z(p^2)$ and
$M_1(\mbox{\boldmath{$p$}}^2)=M_2(\mbox{\boldmath{$p$}}^2)\equiv M(p^2),$
which will also enter in $E_1(\mbox{\boldmath{$p$}})=
E_2(\mbox{\boldmath{$p$}})=E(p)\equiv\sqrt{p^2+M^2(p^2)}.$ The set of coupled
equations governing the radial functions $\Phi_1(p)$ and $\Phi_2(p)$ in our
independent Salpeter components $\Phi_1(\mbox{\boldmath{$p$}})$ and
$\Phi_2(\mbox{\boldmath{$p$}})$ of ${}^1{\rm S}_0$ states, respectively, can
be simply derived by, for instance, ``dressing'' Eq.~(1)~of
Ref.\cite{Lucha00:IBSEm=0} or Eq.~(1) of Ref.\cite{Lucha00:IBSEnzm} by
insertion of the appropriate factors $Z(p^2)$ in all interaction~terms and by
replacement of all constant constituent masses $m$ by the relevant mass
functions~$M(p^2)$:\begin{eqnarray}&&2\,E(p)\,\Phi_2(p)+Z^2(p^2)\int
\limits_0^\infty\frac{{\rm d}q\,q^2}{(2\pi)^2}\,V_0(p,q)\,\Phi_2(q)=M_{\rm
B}\,\Phi_1(p)\ ,\nonumber\\[1ex]&&2\,E(p)\,\Phi_1(p)\label{Eq:RE}\\[1ex]
&&+\,Z^2(p^2)\int\limits_0^\infty\frac{{\rm d}q\,q^2}{(2\pi)^2}
\left[\frac{M(p^2)}{E(p)}\,V_0(p,q)\,\frac{M(q^2)}{E(q)}
+\frac{p}{E(p)}\,V_1(p,q)\,\frac{q}{E(q)}\right]\Phi_1(q)=M_{\rm
B}\,\Phi_2(p)\ .\nonumber\end{eqnarray}The configuration- and momentum-space
representations of any radial function are related by Fourier--Bessel
transformations which involve spherical Bessel functions of the first~kind
$j_n(z)$ ($n=0,\pm 1,\pm2,\dots$)\cite{Abramowitz}; as a kind of reminiscence
of these, the interaction $V(\mbox{\boldmath{$p$}}-\mbox{\boldmath{$q$}})$~in
the kernel $\hat K(\mbox{\boldmath{$p$}}-\mbox{\boldmath{$q$}})$ enters here
in form of some static potential $V(r)$ in
configuration~space:$$V_L(p,q)\equiv 8\pi\int\limits_0^\infty{\rm
d}r\,r^2\,j_L(p\,r)\,j_L(q\,r)\,V(r)\ ,\quad L=0,1,2,\dots\ .$$

The particular structure of the set of equations (\ref{Eq:RE}) allows to find
its solutions\cite{Lucha00:IBSEm=0,Lucha00:IBSE-C4,Lucha00:IBSEnzm,
Lucha01:IBSEIAS}~by inserting one of these relations into the other and
obtaining an eigenvalue equation for~$M_{\rm B}^2$:\begin{eqnarray}M_{\rm
B}^2\,\Phi_2(p)&=&4\,E^2(p)\,\Phi_2(p)+2\,Z^2(p^2)\,E(p)\int\limits_0^\infty
\frac{{\rm d}q\,q^2}{(2\pi)^2}\,V_0(p,q)\,\Phi_2(q)\nonumber\\[1ex]
&+&2\,\frac{Z^2(p^2)}{E(p)}\int\limits_0^\infty\frac{{\rm d}q\,q^2}{(2\pi)^2}
\left[M(p^2)\,M(q^2)\,V_0(p,q)+p\,q\,V_1(p,q)\right]\Phi_2(q)\nonumber\\[1ex]
&+&Z^2(p^2)\int\limits_0^\infty\frac{{\rm d}q\,q^2}{(2\pi)^2}
\left[\frac{M(p^2)}{E(p)}\,V_0(p,q)\,\frac{M(q^2)}{E(q)}
+\frac{p}{E(p)}\,V_1(p,q)\,\frac{q}{E(q)}\right]\nonumber\\[1ex]
&\times&Z^2(q^2)\int\limits_0^\infty\frac{{\rm
d}k\,k^2}{(2\pi)^2}\,V_0(q,k)\,\Phi_2(k)\ .\label{Eq:MB2}\end{eqnarray}By
expansion over the basis for radial functions summarized in
Appendix~\ref{App:GLB} we convert this integral equation to an equivalent
matrix eigenvalue problem which can be diagonalized by standard means. As
noted in Sec.~\ref{Sec:IBSEWEP}, Salpeter's equation corresponds to the
free-propagator approximation $Z(p^2)\cong1$ and $M(p^2)\cong m.$ Thus the
studies reported in Refs.\cite{Lucha00:IBSEm=0,Lucha00:IBSE-C4,
Lucha00:IBSEnzm,Lucha01:IBSEIAS}~could get the matrix, for a large class of
interactions, in algebraic form. Due to the presence~of the true quark
propagator functions $Z(p^2)$ and $M(p^2),$ in general this is no longer
possible~here. Upon construction of one Salpeter component, $\Phi_2(p),$ as
solution of Eq.~(\ref{Eq:MB2}), its companion, $\Phi_1(p),$ follows, for
$M_{\rm B}\ne0,$ immediately from the first of the two coupled equations
(\ref{Eq:RE}). For vanishing bound-state mass, $M_{\rm B}=0,$
Eqs.~(\ref{Eq:RE}) decouple and are thus solved independently.

\section{Linear confinement: results and discussion}\label{Sec:RD}Let us
eventually apply the formalism developed in Secs.~\ref{Sec:IBSEWEP} through
\ref{Sec:PSFAFBS} to a confining~(static) potential of linear shape,
$V(r)=\lambda\,r,$ with slope $\lambda=0.2\;\mbox{GeV}^2.$ For confining
interactions a time-component Lorentz-vector structure of the kernel appears
to be free of all the stability problems encountered by solutions found for a
kernel of Lorentz-scalar structure\cite{Parramore95,Parramore96,Olsson95}.

\begin{figure}[p]\begin{center}\psfig{figure=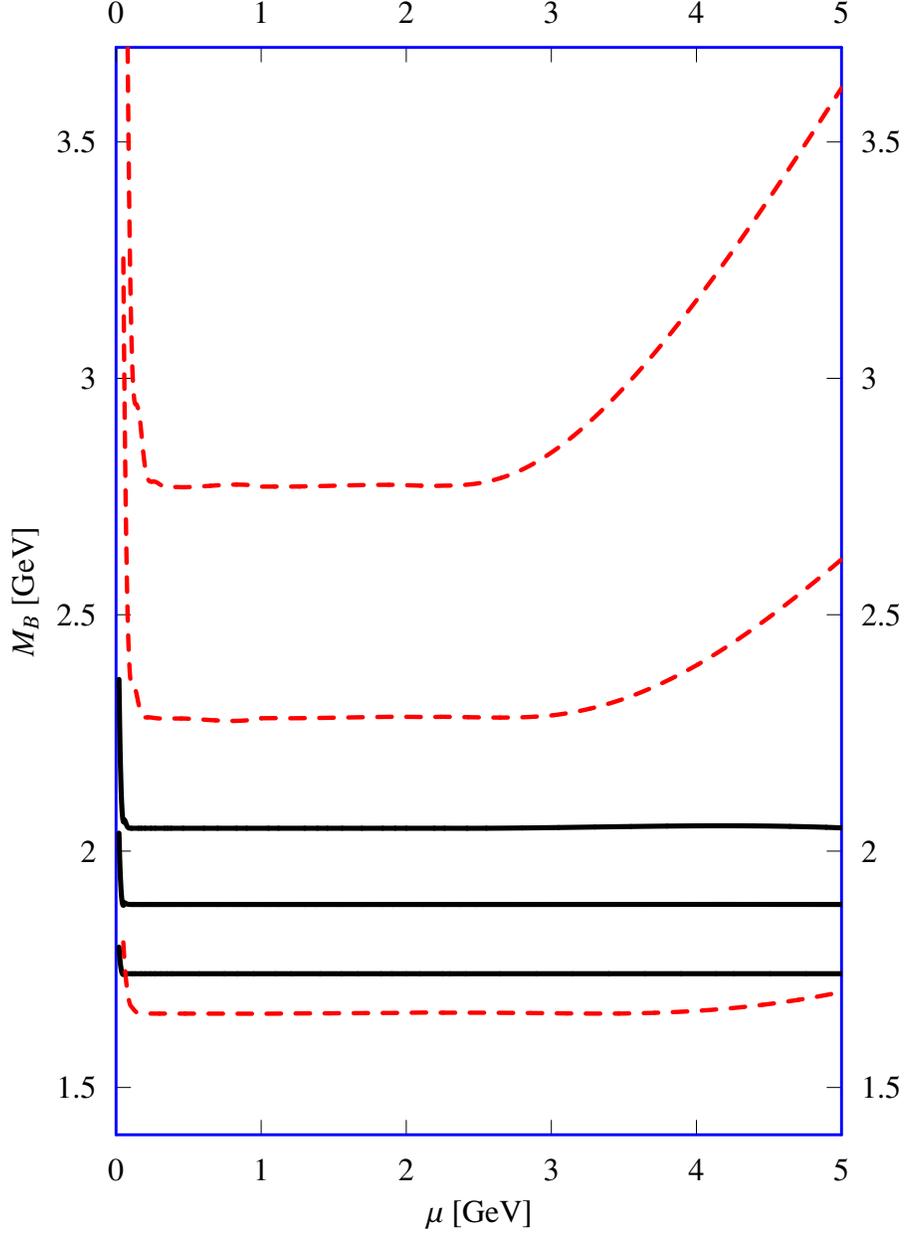,scale=1}
\caption{Bound-state masses $M_{\rm B}$ (as functions of our variational
parameter $\mu$) of the three lowest-lying (positive-norm) $J^{PC}=0^{-+}$
eigenstates of our exact-propagator instantaneous Bethe--Salpeter equation
(\ref{Eq:IBSE}) for the propagator parametrization (\ref{Eq:ProPar}) but
$m_0=0$ ({\sl full~lines\/}) and of the Salpeter equation for massless
constituents ({\sl dashed lines\/}), for a time-component Lorentz-vector
kernel representing some linear potential $V(r)=\lambda\,r$ of slope
$\lambda=0.2\;\mbox{GeV}^2.$ These results arise from diagonalization, for
given values of $\mu,$ of $50\times50$ matrices equivalent to our
exact-propagator equation but, mimicking Ref.\cite{Lucha00:IBSEm=0}, only
$15\times15$ matrices equivalent to Salpeter's equation and in both cases 50
terms in all intermediate-step series expansions.}\label{Fig:MBCL}
\end{center}\end{figure}

Our first goal is to analyze the effect of the dynamical generation of quark
masses on the solutions of the instantaneous Bethe--Salpeter equation
(\ref{Eq:IBSE}) with exact propagators. To this end, Fig.~\ref{Fig:MBCL}
compares, for the three lowest positive-norm $J^{PC}=0^{-+}$ bound states,
the mass eigenvalues $M_{\rm B}$ of Eq.~(\ref{Eq:IBSE}) for exact light-quark
propagators with $m_0=0$ (corresponding to the chiral limit of QCD) with
those of a Salpeter equation for massless constituents
\cite{Lucha00:IBSEm=0,Lucha00:IBSE-C4}.

The chosen bases for the Hilbert space $L_2(R^+)$ of all with the weight
function $w(r)=r^2$ square-integrable (``radial'') functions on the positive
real line $R^+$ introduce one additional degree of freedom, by allowing the
basis states to depend on a variational parameter $\mu>0.$ As a basis, these
vectors constitute a complete orthonormal system for any particular~value of
$\mu.$ Therefore, as long as relying exclusively on expansions over the {\em
full\/} set of basis vectors our results may be expected to be independent of
$\mu.$ Necessary truncations of expansions to a finite number of basis
vectors will induce a certain amount of $\mu$-dependence of the~results.
However, a reasonable technique involving expansions should exhibit stability
with respect to the increase of the number of basis vectors. If taking into
account a large enough number of basis vectors, by reducing the dependence on
$\mu$ some ``region of stability'' should emerge.

With respect to the mass $M_{\rm B}$ of a given bound state, such a region of
stability manifests itself in form of a plateau where the numerical value
predicted for $M_{\rm B}$ is constant over some nonvanishing range of $\mu.$
Beyond doubt, the formation of these plateaus is obvious in
Fig.~\ref{Fig:MBCL}. Such extrema of $M_{\rm B}$ disclose the bound states.
Compared with the free-propagator~results, the ground state of
Eq.~(\ref{Eq:IBSE}) is higher but its radial excitations are lower, to the
effect~that~all level spacings are significantly smaller if using exact
propagators. In general, it is, of~course, not possible to compensate for
these shifts by some change of the parameter values entering in the
interaction kernel. For instance, a reduction of the slope $\lambda$ of the
linear potential~by~a factor 2 to $\lambda\cong0.1\;\mbox{GeV}^2$ lowers the
ground-state energy eigenvalue of Eq.~(\ref{Eq:IBSE}) to the level of its
Salpeter-equation counterpart but simultaneously diminishes the level
spacings further.

Table~\ref{Tab:MB-EPFP} lists the masses $M_{\rm B}$ of the three
lowest-lying (positive-norm) $J^{PC}=0^{-+}$ bound states calculated from the
exact-propagator instantaneous Bethe--Salpeter equation (\ref{Eq:IBSE}) for
the full ($m_0\ne0$) parametrization (\ref{Eq:ProPar}) and (\ref{Eq:ParVal})
of the exact light-quark propagators. Because of the relative smallness of
the (explicitly chiral-symmetry breaking) current mass $m_0$ these
eigenvalues are, of course, less than some 0.5\% larger than the
corresponding ``chiral-limit'' values forming the stability plateaus
discernible in Fig.~\ref{Fig:MBCL}. These mass values are confronted in
Table~\ref{Tab:MB-EPFP} with the corresponding mass eigenvalues $M_{\rm B}$
of the Salpeter equation, computed by assuming, for the constituent mass $m$
of the light {\sl u}- and {\sl d}-quarks entering their effective
propagators, the typical value of $m=0.336\;{\rm GeV}$ frequently adopted by
nonrelativistic and relativistic constituent quark models to describe hadrons
as bound states of quarks\cite{Lucha91,Lucha92}. Raising in the Salpeter
equation the constituent quark mass $m$ from $m=0$ to the canonical value
$m=0.336\;{\rm GeV}$ shifts the masses $M_{\rm B}$ of the three lowest states
by more than $0.1\;{\rm GeV}$ towards larger values. The net result of this
is a further increase of the discrepancy between the level spacings predicted
by the exact-propagator equation (\ref{Eq:IBSE}) and the ones arising~from
the free-propagator Salpeter equation with some kind of appropriate or
reasonable effective mass of the constituent quarks. Therefore, we are forced
to conclude that any neglect of the proper behaviour of the
momentum-dependent quark mass, $M(p^2),$ by approximating it by a constant
constituent mass $m$ is, at least for the light {\sl u}- and {\sl d}-quarks,
rather~questionable.

\begin{table}[ht]\caption{Bound-state masses $M_{\rm B}$ (in units of GeV)
for the three lowest-lying positive-norm $J^{PC}=0^{-+}$ eigenstates (denoted
by $1^1{\rm S}_0,$ $2^1{\rm S}_0,$ $3^1{\rm S}_0,$ in usual spectroscopic
notation) of our exact-propagator instantaneous Bethe--Salpeter equation
(\ref{Eq:IBSE}) with light-quark propagators parametrized by
Eqs.~(\ref{Eq:ProPar}), (\ref{Eq:ParVal}) and of the free-propagator Salpeter
equation with light-quark constituent mass $m=0.336\;{\rm GeV},$ for
time-component Lorentz-vector kernels representing a linear potential
$V(r)=\lambda\,r$ of slope $\lambda=0.2\;\mbox{GeV}^2,$ obtained by
converting both equations to $50\times50$ matrices and taking into account 50
terms in intermediate-step series expansions.}\label{Tab:MB-EPFP}
\begin{center}\begin{tabular}{ccc}\hline\hline&&\\[-1.5ex]\multicolumn{1}{c}
{State}&\multicolumn{1}{c}{\begin{tabular}{c}Exact-propagator\\bound-state
equation\end{tabular}}&\multicolumn{1}{c}{\begin{tabular}{c}(Free-propagator)\\
Salpeter equation\end{tabular}}\\[2.5ex]\hline\\[-1.5ex] $1^1{\rm
S}_0$&1.750&1.813\\$2^1{\rm S}_0$&1.895&2.410\\$3^1{\rm
S}_0$&2.056&2.889\\[1ex]\hline\hline\end{tabular}\end{center}\end{table}

\begin{figure}[p]\begin{center}\begin{tabular}{c}
\psfig{figure=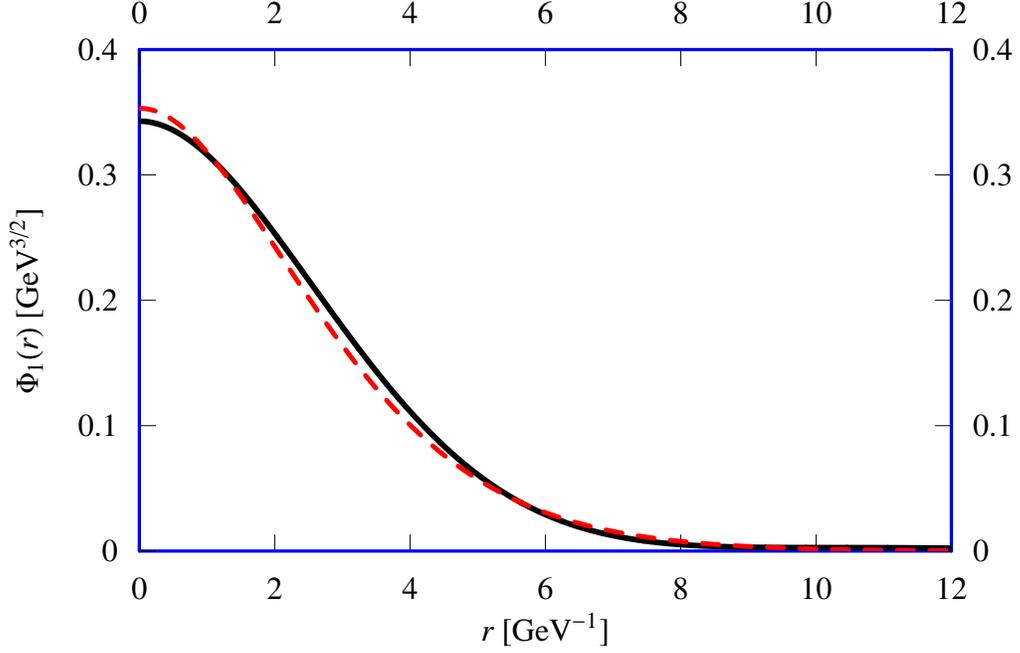,scale=1}\\[0.05ex](a)\end{tabular}\\[1.5ex]
\begin{tabular}{c}\psfig{figure=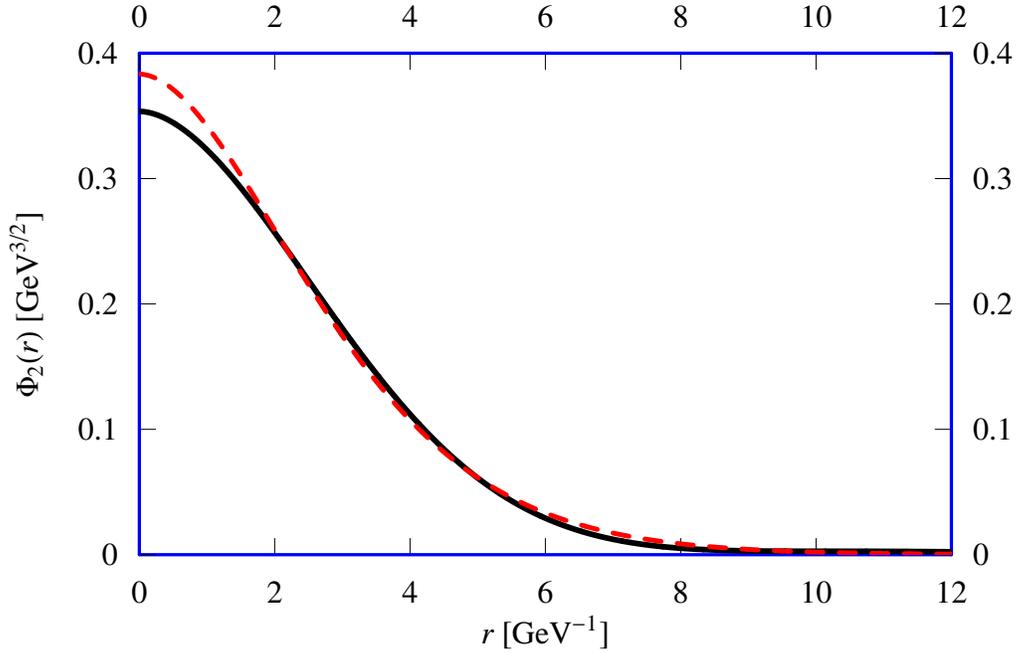,scale=1}\\[0.05ex](b)
\end{tabular}\\[0.5ex]\caption{Configuration-space radial Salpeter component
functions $\Phi_1(r)$ (a) and $\Phi_2(r)$ (b) for the lowest positive-norm
$J^{PC}=0^{-+}$ bound state of the exact-propagator instantaneous
Bethe--Salpeter equation (\ref{Eq:IBSE}) with the quark propagator
parametrization (\ref{Eq:ProPar}) ({\sl full lines\/}) and of the Salpeter
equation with a light-quark {\em constituent\/} mass $m=0.336\;{\rm GeV}$
({\sl dashed lines\/}) for time-component Lorentz-vector kernels with linear
potential $V(r)=\lambda\,r,$ $\lambda=0.2\;\mbox{GeV}^2.$}\label{Fig:Phi12}
\end{center}\end{figure}

Figure~\ref{Fig:Phi12} illustrates for the $J^{PC}=0^{-+}$ ground state
($1^1{\rm S}_0$ state) of the exact-propagator instantaneous Bethe--Salpeter
equation (\ref{Eq:IBSE}) with quark-propagator parametrization
(\ref{Eq:ProPar}) the configuration-space behaviour of the radial Salpeter
component functions $\Phi_1(r)$ and $\Phi_2(r).$ The norm $\|\Phi\|$ of the
Salpeter amplitude $\Phi(\mbox{\boldmath{$p$}})$ for $J^{PC}=0^{-+}$ bound
states reads\cite{Lagae92a,Olsson95,Lucha00:IBSEm=0}
$$\|\Phi\|^2=4\int\frac{{\rm d}^3p}{(2\pi)^3}\,
[\Phi_1^\ast(\mbox{\boldmath{$p$}})\,\Phi_2(\mbox{\boldmath{$p$}})
+\Phi_2^\ast(\mbox{\boldmath{$p$}})\,\Phi_1(\mbox{\boldmath{$p$}})]\ ;$$this
translates for the radial parts $\Phi_1(p),$ $\Phi_2(p)$ of the Salpeter
components $\Phi_1(\mbox{\boldmath{$p$}}),$
$\Phi_2(\mbox{\boldmath{$p$}})$~to
$$\|\Phi\|^2=\frac{4}{(2\pi)^3}\int\limits_0^\infty{\rm
d}p\,p^2\,[\Phi_1^\ast(p)\,\Phi_2(p)+\Phi_2^\ast(p)\,\Phi_1(p)]\ .$$For the
plots depicted in Fig.~\ref{Fig:Phi12}, the normalization has been chosen
such that $\Phi_2(r)$ satisfies$$\int\limits_0^\infty{\rm
d}r\,r^2\,|\Phi_2(r)|^2=\int\limits_0^\infty{\rm d}p\,p^2\,|\Phi_2(p)|^2=1\
;$$the first of Eqs.~(\ref{Eq:RE}) fixes, after a Fourier--Bessel
transformation, the normalization of~$\Phi_1(r).$ A normalization factor
common to $\Phi_1(p)$ and $\Phi_2(p)$ will then give any desired value
to~$\|\Phi\|.$

A closer inspection of the radial Salpeter components $\Phi_1(r)$ and
$\Phi_2(r)$ reveals a striking similarity to their counterparts found as
solutions of the Salpeter equation for a constituent quark mass of
$m=0.336\;{\rm GeV}.$ Exact- and free-propagator Salpeter components
$\Phi_1(r)$ and $\Phi_2(r)$ show some notable difference only for $r<5\;{\rm
GeV}^{-1}$ for $\Phi_1(r)$ and for $r<2\;{\rm GeV}^{-1}$~for $\Phi_2(r).$
They are hardly distinguishable from each other for $r>5\;{\rm GeV}^{-1}$ for
both $\Phi_1(r)$ and $\Phi_2(r).$ Thus we feel entitled to expect similar
predictions for quantities such as decay~rates.

\section{Summary, conclusions, and outlook}\label{Sec:SCO}The reduction of
the Bethe--Salpeter formalism to the Salpeter equation requires to assume for
all bound-state constituents not only an instantaneous interaction but also
free-particle propagation. Realizing this fact, a bound-state equation that
retains the exact propagators of the bound-state constituents and generalizes
Salpeter's equation has been formulated~by consequent application of the
instantaneous approximation to all exact propagators
too\cite{Lucha05:IBSEWEP}. Of course, this may be extended to Bethe--Salpeter
equations for bound states composed of particles that are not, or not all,
identical to spin-$\frac{1}{2}$ fermions as well as to three-dimensional
reductions\cite{Babutsidze98,Babutsidze99,Kopaleishvili01,Babutsidze03} of
the Bethe--Salpeter equation\cite{BSE} different from Salpeter's
equation\cite{SE}.

The present investigation addressed the question of how realistic
descriptions of mesons as quark--antiquark bound states within a general
instantaneous Bethe--Salpeter formalism will be modified by such
reinstallment of the exact quark propagators, extracted from QCD by analytic
continuation from Euclidean to Minkowski space. Interestingly, for the
example of a specific interaction used already in earlier studies
\cite{Olsson95, Olsson96,Lucha00:IBSEm=0,Lucha00:IBSE-C4,Lucha00:IBSEnzm,
Lucha01:IBSEIAS}, we find a drastic shrinking of the level spacings of the
bound states while their amplitudes remain practically unchanged.

\section*{Acknowledgements}One of us (W.~L.) would like to thank Craig
D.~Roberts for a lot of stimulating discussions, as well as for communicating
his parametrizations of the (numerical) propagator functions.

\appendix\section{The generalized Laguerre basis for radial functions}
\label{App:GLB}The Hilbert space $L_2(R^3)$ of all square-integrable
functions on the 3-dimensional Euclidean space $R^3$ can be spanned by basis
functions each of which is the product of a function of the radial variable
and of an angular term, the latter being represented by a spherical harmonic
${\cal Y}_{\ell m}(\Omega)$ for the angular momentum $\ell=0,1,2,\dots$ and
its projection $m=-\ell,-\ell+1,\dots,+\ell$ which all depend on the solid
angle $\Omega\equiv(\theta,\phi)$ and satisfy the orthonormalization
condition$$\int{\rm d}\Omega\,{\cal Y}^\ast_{\ell m}(\Omega)\,{\cal
Y}_{\ell'm'}(\Omega)=\delta_{\ell\ell'}\,\delta_{mm'}\ .$$For each value of
$\ell$ the radial functions constitute a basis for the Hilbert space
$L_2(R^+)$ of all with the weight function $w(r)=r^2$ square-integrable
functions on the positive real line $R^+.$ The basis functions of $L_2(R^3)$
in configuration and momentum space are related by Fourier transformation.
Thus, the configuration-space representation $\phi_i^{(\ell)}(r)$ and
momentum-space representation $\phi_i^{(\ell)}(p)$ of the radial factors are
related by the Fourier--Bessel transformation
\begin{eqnarray*}\phi_i^{(\ell)}(r)&=&{\rm
i}^\ell\,\sqrt{\frac{2}{\pi}}\int\limits_0^\infty{\rm
d}p\,p^2\,j_\ell(p\,r)\,\phi_i^{(\ell)}(p)\ ,\quad i=0,1,2,\dots\
,\quad\ell=0,1,2,\dots\ ,\\[1ex]\phi_i^{(\ell)}(p)&=&(-{\rm
i})^\ell\,\sqrt{\frac{2}{\pi}}\int\limits_0^\infty{\rm
d}r\,r^2\,j_\ell(p\,r)\,\phi_i^{(\ell)}(r)\ ,\quad i=0,1,2,\dots\
,\quad\ell=0,1,2,\dots\ .\end{eqnarray*}The spherical Bessel functions of the
first kind, $j_n(z)$ ($n=0,\pm 1,\pm 2,\dots$)\cite{Abramowitz}, are remnants
of the angular integration. This may be easily deduced with the help of the
expansion~of the plane waves over spherical harmonics ${\cal Y}_{\ell m}$ in
configuration $(\Omega_r)$ and momentum $(\Omega_p)$~space$$\exp({\rm
i}\,\mbox{\boldmath{$p$}}\cdot\mbox{\boldmath{$x$}})=4\pi\,\sum_{\ell=0}^\infty
\,\sum_{m=-\ell}^{+\ell}\,{\rm i}^\ell\,j_\ell(p\,r)\,{\cal Y}^\ast_{\ell
m}(\Omega_p)\,{\cal Y}_{\ell m}(\Omega_r)\ .$$

In terms of orthogonal polynomials of generalized-Laguerre type (for
parameter $\gamma$)\cite{Abramowitz},$$L_i^{(\gamma)}(x)=\sum_{t=0}^i\,
(-1)^t\left(\begin{array}{c}i+\gamma\\i-t\end{array}\right)\frac{x^t}{t!}\
,\quad i=0,1,2,\dots\ ,$$which, by construction, are orthonormalized [with
weight $w(x)=x^\gamma\exp(-x)]$ according~to$$\int\limits_0^\infty{\rm
d}x\,x^\gamma\exp(-x)\,L_i^{(\gamma)}(x)\,L_j^{(\gamma)}(x)=
\frac{\Gamma(\gamma+i+1)}{i!}\,\delta_{ij}\ ,\quad i,j=0,1,2,\dots\ ,$$our
favourite choice of these radial bases is defined in configuration-space
representation~by$$\phi_i^{(\ell)}(r)=\sqrt{\frac{(2\,\mu)^{2\,\ell+3}\,i!}
{\Gamma(2\,\ell+i+3)}}\,r^\ell\exp(-\mu\,r)\, L_i^{(2\,\ell+2)}(2\,\mu\,r)\
,\quad i=0,1,2,\dots\ .$$These basis functions involve one positive real
variational parameter, with the dimension of mass, $\mu.$ The requirement of
their normalizability imposes the constraint $\mu>0.$ Then these
configuration-space radial basis functions, $\phi_i^{(\ell)}(r),$ satisfy the
orthonormalization~condition$$\int\limits_0^\infty{\rm d}r\,r^2\,
\phi_i^{(\ell)}(r)\,\phi_j^{(\ell)}(r)=\delta_{ij}\ ,\quad i,j=0,1,2,\dots\
.$$Note that the configuration-space representation of our basis functions is
chosen to be~real.

\newpage

The corresponding momentum-space representation $\phi_i^{(\ell)}(p)$ of our
basis functions reads\begin{eqnarray*}\phi_i^{(\ell)}(p)
&=&\sqrt{\frac{(2\,\mu)^{2\,\ell+3}\,i!}{\Gamma(2\,\ell+i+3)}}\,\frac{(-{\rm
i})^\ell\,p^\ell}{2^{\ell+1/2}\,\Gamma\left(\ell+\frac{3}{2}\right)}\,\\[1ex]
&\times&\sum_{t=0}^i\,\frac{(-1)^t}{t!}\left(\begin{array}{c}i+2\,\ell+2\\
i-t\end{array}\right)\frac{\Gamma(2\,\ell+t+3)\,(2\,\mu)^t}
{(p^2+\mu^2)^{(2\,\ell+t+3)/2}}\\[1ex]
&\times&F\left(\frac{2\,\ell+t+3}{2},-\frac{1+t}{2};\ell+\frac{3}{2};
\frac{p^2}{p^2+\mu^2}\right),\quad i=0,1,2,\dots\ ,\end{eqnarray*} with the
hypergeometric series $F(u,v;w;z)$ given in terms of the gamma function
$\Gamma$ by\cite{Abramowitz}
$$F(u,v;w;z)=\frac{\Gamma(w)}{\Gamma(u)\,\Gamma(v)}\,\sum_{n=0}^\infty\,
\frac{\Gamma(u+n)\,\Gamma(v+n)}{\Gamma(w+n)}\,\frac{z^n}{n!}\ .$$The
momentum-space radial basis functions $\phi_i^{(\ell)}(p)$ fulfill the
orthonormalization condition$$\int\limits_0^\infty{\rm
d}p\,p^2\,\phi_i^{\ast(\ell)}(p)\,\phi_j^{(\ell)}(p)=\delta_{ij}\ ,\quad
i,j=0,1,2,\dots\ .$$In momentum space, our basis functions are real for
$\ell=0,$ as well as for all even values~of~$\ell$:$$\phi_i^{\ast(\ell)}(p)=
\phi_i^{(\ell)}(p)\quad\mbox{for\ }\ell=0,2,4,\dots\ ,\quad\forall\
i=0,1,2,\dots\ .$$The virtue of our bases is their analytic availability in
configuration {\em and\/} momentum space.

Mainly for computational convenience, the present investigation makes use of
the radial basis functions for two values $\ell=0$ and $\ell=1$ of the
angular momentum. Having to deal, in momentum-space representation, with the
cumbersome hypergeometric series~$F(u,v;w;z)$ may be avoided by employing
simplified expressions equivalent to the above definition\cite{Lucha97}:
\begin{eqnarray*}
\phi_i^{(0)}(p)&=&\sqrt{\frac{i!}{\mu\,\pi\,\Gamma(i+3)}}\,\frac{4}{p}\,
\sum_{t=0}^i\,(-2)^t\,(t+1)\left(\begin{array}{c}i+2\\i-t\end{array}\right)
\left(1+\frac{p^2}{\mu^2}\right)^{-(t+2)/2}\\[1ex]
&\times&\sin\left((t+2)\arctan\frac{p}{\mu}\right)\\[1ex] &=&\frac{{\rm
Im}\{(p+{\rm i}\,\mu)^{2\,i+3}\,[p-{\rm i}\,(3+2\,i)\,\mu]\}}
{\sqrt{\mu\,\pi\,(i+1)\,(i+2)}\,p\,(p^2+\mu^2)^{2+i}}\
,\\[1ex]\phi_i^{(1)}(p) &=&-{\rm
i}\,\sqrt{\frac{\mu^5}{\pi\,(i+1)\,(i+2)\,(i+3)\,(i+4)}}\,
\frac{8}{p^2}\,\sum_{t=0}^i\,\frac{(-2)^t}{t!}
\left(\begin{array}{c}i+4\\i-t\end{array}\right)
\frac{(t+3)!\,\mu^t}{(p^2+\mu^2)^{(t+3)/2}}\\[1ex]&\times&
\left[\frac{\sqrt{p^2+\mu^2}}{t+2}\sin\left((t+2)\arctan\frac{p}{\mu}\right)-
\frac{\mu}{t+3}\sin\left((t+3)\arctan\frac{p}{\mu}\right)\right]\\[1ex]
&=&\frac{{\rm i}}{2\,\sqrt{\mu^3\,\pi\,(i+1)\,(i+2)\,(i+3)\,(i+4)}\,p^2\,
(p^2+\mu^2)^3}\\[1ex]&\times& {\rm Im}\left\{\frac{(p-{\rm
i}\,\mu)^{i+5}}{(p+{\rm i}\,\mu)^i}\, [3\,p^3+3\,{\rm
i}\,(5+2\,i)\,p^2\,\mu-(5+2\,i)^2\,p\,\mu^2-{\rm i}\,
(5+2\,i)\,\mu^3]\right\}.\end{eqnarray*}

\end{document}